\documentclass[a4paper,11pt]{article}
\usepackage[T2A]{fontenc}
\usepackage[utf8]{inputenc}
\usepackage[english]{babel}
\usepackage{amsmath}
\usepackage{amsfonts}
\usepackage{amssymb}
\usepackage{wasysym}
\usepackage{amsthm}
\usepackage{tikz}
\usepackage{tikz-cd}
\usetikzlibrary{decorations.markings}
\usetikzlibrary{positioning}
\usepackage{braids}
\usepackage{array}
\usepackage{ytableau}
\usepackage{longtable}
\usepackage{empheq}
\usepackage{multicol}
\usepackage{mathrsfs}
\usepackage{amssymb}
\usepackage{diagbox}
\usepackage{color}
\usepackage{physics}
\usepackage{ytableau}
\usepackage{cite}
\definecolor{lgreen}{rgb}{0.9,1,0.8}

\usepackage[most]{tcolorbox}
\tcbset{boxrule=0.6pt,colback=white!97!green}

\usepackage{pdflscape}
\usepackage{array, longtable}
\newcolumntype{C}{>{$}c<{$}}

\usepackage{pb-diagram}
\usepackage{setspace}
\usepackage[hidelinks,colorlinks=true,unicode]{hyperref}
\hypersetup{
	linkcolor=blue,
	citecolor=blue,
	filecolor=magenta,
	urlcolor=blue,
}
\usepackage{url}
\usepackage[left=2.3cm,right=2.3cm,top=2.3cm,bottom=2.3cm,bindingoffset=0cm]{geometry}

\def\p{\partial}

\def\myblue{white!40!blue}

\def\uhbox{\begin{tikzpicture}[scale=0.15]
		\foreach \i/\j in {0/0}
		{
			\draw[thick] (\i,\j) -- (\i+1,\j);
		}
		\foreach \i/\j in {0/0}
		{
			\draw[thick] (\i,\j) -- (\i,\j-1);
		}
		\foreach \i/\j in {1/0}
		{
			\draw[thick] (\i,\j) -- (\i-1,\j-1);
		}
\end{tikzpicture}}
\def\lhbox{\begin{tikzpicture}[scale=0.15]
		\foreach \i/\j in {0/-1}
		{
			\draw[thick] (\i,\j) -- (\i+1,\j);
		}
		\foreach \i/\j in {1/0}
		{
			\draw[thick] (\i,\j) -- (\i,\j-1);
		}
		\foreach \i/\j in {1/0}
		{
			\draw[thick] (\i,\j) -- (\i-1,\j-1);
		}
\end{tikzpicture}}
\def\fullbox{\begin{tikzpicture}[scale=0.15]
		\foreach \i/\j in {0/0, 0/-1}
		{
			\draw[thick] (\i,\j) -- (\i+1,\j);
		}
		\foreach \i/\j in {0/0,1/0}
		{
			\draw[thick] (\i,\j) -- (\i,\j-1);
		}
		\foreach \i/\j in {}
		{
			\draw[thick] (\i,\j) -- (\i-1,\j-1);
		}
\end{tikzpicture}}

\def\shtile{\begin{tikzpicture}[scale=0.15]
		\draw (0,0) -- (1,0) -- (0,-1) -- cycle;
\end{tikzpicture}}
\def\shhtile{\begin{tikzpicture}[scale=-0.15]
		\draw (0,0) -- (1,0) -- (0,-1) -- cycle;
\end{tikzpicture}}

\definecolor{red}{rgb}{1,0,0}
\definecolor{orange}{rgb}{1,0.5,0}
\definecolor{violet}{rgb}{0.7,0,1}

\def\be{\begin{eqnarray}}
	\def\ee{\end{eqnarray}}

\begin{document}
	\hfill MIPT/TH-02/25
	
	\hfill ITEP/TH-02/25
	
	\hfill IITP/TH-02/25
	
	\vskip 1.5in
	%\vskip 1cm
	\begin{center}
		
		{\bf\Large Super-Hamiltonians for super-Macdonald polynomials}
		
		\vskip 0.2in
		\renewcommand{\thefootnote}{\fnsymbol{footnote}}
		{Dmitry Galakhov$^{1,2,3}$\footnote[2]{e-mail: galakhov@itep.ru},  Alexei Morozov$^{1,2,3}$\footnote[3]{e-mail: morozov@itep.ru} and Nikita Tselousov$^{1,2,3}$\footnote[4]{e-mail: tselousov.ns@phystech.edu}}\\
		\vskip 0.2in
		\renewcommand{\thefootnote}{\roman{footnote}}
		{\small{
				\textit{$^1$MIPT, 141701, Dolgoprudny, Russia}
				\vskip 0 cm
				\textit{$^2$NRC “Kurchatov Institute”, 123182, Moscow, Russia}
				\vskip 0 cm
				\textit{$^3$IITP RAS, 127051, Moscow, Russia}
				\vskip 0 cm
				\textit{$^4$ITEP, Moscow, Russia}
		}}
	\end{center}
	
	\vskip 0.2in
	\baselineskip 16pt
	
	\centerline{ABSTRACT}
	
	\bigskip
	
	{\footnotesize
		The Macdonald finite-difference Hamiltonian is lifted to a super-generalization.
		In addition to canonical bosonic time variables $p_k$ new Grassmann time variables $\theta_k$ are introduced, and the Hamiltonian is represented as a differential operator acting on a space of functions of both types of variables $p_k$ and $\theta_k$.
		Eigenfunctions for this Hamiltonian are a suitable generalization of Macdonald polynomials to super-Macdonald polynomials discussed earlier in the literature.
		Peculiarities of the construction in comparison to the canonical bosonic case are discussed.
	}

	\bigskip
	
	\section{Introduction}
	
	The extension of Macdonald polynomials \cite{Macdonald} to the super-case \cite{GMT2, GMT3, Blondeau-Fournier:2011sft, Blondeau-Fournier:2012exj} appears to be applicable to studies of super-Yangians like $Y(\widehat {\mathfrak{gl}}_{1,1})$ \cite{Galakhov:2023mak, Galakhov:2024foa, GGMT} and their super-DIM generalizations \cite{GMT3, Noshita:2021dgj, Galakhov:2021vbo}.
	Surely, a modification (introduction of Grassmann variables) of the original Macdonald construction sheds some light on peculiarities of the original construction itself: we might try to question whether an appearance of algebraic and combinatorial structures accompanying Macdonald polynomials is accidental and may be naturally extended to other similar constructions.
	
	In approaches used in \cite{GMT2, GMT3, Blondeau-Fournier:2011sft} the super-generalization was based on a variety of intuitive arguments.
	among which the major role was played by a generalization of Cauchy formulas and triangularity property. This is the old idea behind the definition of the far more general Kerov functions \cite{kerov1991hall, Mironov:2020aaa, Mironov:2019uaa, Mironov:2019exq}, where Macdonalds are distinguished by the independence of the ordering of Young diagrams -- which makes triangularity a well-defined notion.
	
	However, the most conventional approach defines Macdonald polynomials as eigenfunctions
	of Ruijsenaars Hamiltonians \cite{ruijsenaars1987complete,RUIJSENAARS1986, Mironov:2019uoy}.
	In this paper we also exploit this idea and construct a set of commuting super-Hamiltonians whose eigenfunctions turn out to coincide with canonically defined super-Macdonald polynomials. We discuss special properties of these Hamiltonians that are relevant for the representation theories of infinite dimensional algebras -- such as Yangians \cite{Prochazka:2015deb, Tsymbaliuk2017, Galakhov:2024mbz,Morozov:2023vra, Morozov:2022ndt} and DIM \cite{Ding:1996mq,Miki:2007mer,Awata:2017lqa, Mironov:2024sbc, Feigin2, Awata:2018svb} algebras where these operators play a role of Cartan generators. Super-Hamiltonian approach to super-Macdonald polynomials was also considered in \cite{Blondeau-Fournier:2012exj, Blondeau-Fournier:2014lba}. Another approach to Macdonald polynomials recently discussed in \cite{Mironov:2024yzw}. Supersymmetric Hamiltonians with extended supersymmetry were constructed in papers \cite{Krivonos:2019aqn,Kozyrev:2021enj, Krivonos:2022jwf}.

	\bigskip
	
	The paper consists of three parts.
	
	Section \ref{Schur} is a brief reminder of the well-known Hamiltonians for Schur polynomials and a discussion of the "sum over boxes" property for eigenvalues and its connection to Pierri rules. In Sections \ref{sSchur} and \ref{Macs} we provide basic facts about super-Schur and Macdonald cases. Section \ref{sMac} is devoted to super-generalizations of Macdonald Hamiltonians --
	the main part of this paper. Section \ref{conc} contains a short conclusion and discussion of the future problems.
	
	\section{Hamiltonians for Schur polynomials
		\label{Schur}}
	\subsection{The simplest Hamiltonians}
	The grading (dilatation) operator is the simplest Hamiltonian for Schur polynomials:
	\begin{equation}
		\hat {\text{G}}_{\fullbox} = \sum_{k=1}^\infty k \, p_k\frac{\p}{\p p_k}
	\end{equation}
	Its eigenvalues count the number of boxes in the Young diagram:
	\begin{equation}
		\hat{\text{G}}_{\fullbox} \, S_{\lambda} = \left( \sum_{\fullbox \in \lambda} \, 1 \right) \cdot  S_{\lambda} = |\lambda| \cdot S_{\lambda}
	\end{equation}
	there is a whole tower of $W$-operators starting with the   cut-and-join operator \cite{MMN,Mironov:2019mah}:
	\begin{equation}
		\hat W = \frac{1}{2}\sum_{a,b=1}^\infty \left(ab \cdot p_{a+b} \frac{\p^2}{\p p_a\p p_b} + (a+b) \cdot p_ap_b\frac{\p}{\p p_{a+b}}\right)
	\end{equation}
	with the eigenvalues $\kappa_{\lambda}$:
	\begin{equation}
		\label{eigenval cut-and-join}
		\hat W \, S_{\lambda} = \kappa_{\lambda} \cdot S_{\lambda} \hspace{20mm}  \boxed{\kappa_{\lambda} = \sum_{\fullbox \in \lambda} (j_{\fullbox} - i_{\fullbox})}
	\end{equation}
	
	Note that the eigenvalues are given by the \textit{sum over boxes} in the Young diagram. 
	We denote by $i_{\Box}$ and $j_{\Box}$ the vertical and horizontal coordinates respectively of a box $\Box$ in the Young diagram. 
	Grading operators are represented as sums of units over the boxes, and higher Hamiltonians are sums of coordinate dependent functions like $\omega_{\fullbox}$. 
	``Sum over boxes'' property of Hamiltonians will be important in construction of higher Hamiltonians. \\
	
	Another relevant for us property of Schur polynomials is the Pierri rules:
	\begin{equation}
		\label{Pierri rule 1}
		\boxed{
			p_1 \cdot S_{\lambda} = \sum_{\fullbox \in \text{Add}(\lambda)} S_{\lambda + \fullbox}
		}	
	\end{equation}
	\begin{equation}
		\label{Pierri rule 2}
		\boxed{
			\frac{\p }{\p p_1 } \cdot S_{\lambda} = \sum_{\fullbox \in \text{Rem}(\lambda)} S_{\lambda - \fullbox}
		}
	\end{equation}
	
	\begin{figure}[ht!]
		\centering
		\begin{tikzpicture}[scale=0.3]
			\foreach \x/\y/\z/\w in {0/0/1/0, 0/-1/1/-1, 0/0/0/-1, 1/0/1/-1, 0/-2/1/-2, 0/-1/0/-2, 1/-1/1/-2, 0/-3/1/-3, 0/-2/0/-3, 1/-2/1/-3, 0/-4/1/-4, 0/-3/0/-4, 1/-3/1/-4, 0/-5/1/-5, 0/-4/0/-5, 1/-4/1/-5, 0/-6/1/-6, 0/-5/0/-6, 1/-5/1/-6, 1/0/2/0, 1/-1/2/-1, 2/0/2/-1, 1/-2/2/-2, 2/-1/2/-2, 1/-3/2/-3, 2/-2/2/-3, 1/-4/2/-4, 2/-3/2/-4, 1/-5/2/-5, 2/-4/2/-5,  2/0/3/0, 2/-1/3/-1, 3/0/3/-1, 2/-2/3/-2, 3/-1/3/-2, 2/-3/3/-3, 3/-2/3/-3, 2/-4/3/-4, 3/-3/3/-4, 3/0/4/0, 3/-1/4/-1, 4/0/4/-1, 3/-2/4/-2, 4/-1/4/-2, 3/-3/4/-3, 4/-2/4/-3, 3/-4/4/-4, 4/-3/4/-4, 4/0/5/0, 4/-1/5/-1, 5/0/5/-1, 4/-2/5/-2, 5/-1/5/-2, 4/-3/5/-3, 5/-2/5/-3, 5/0/6/0, 5/-1/6/-1, 6/0/6/-1, 5/-2/6/-2, 6/-1/6/-2, 5/-3/6/-3, 6/-2/6/-3, 6/0/7/0, 6/-1/7/-1, 7/0/7/-1, 6/-2/7/-2, 7/-1/7/-2, 7/0/8/0, 7/-1/8/-1, 8/0/8/-1, 8/0/9/0, 8/-1/9/-1, 9/0/9/-1}
			{
				\draw (\x,\y) -- (\z,\w);
			}
			\draw[-stealth] (0,0) -- (0,-7);
			\draw[-stealth] (0,0) -- (10,0);
			\node[left] at (0,-7) {$\scriptstyle i$};
			\node[right] at (10,0) {$\scriptstyle j$};
			\node at (4,-8) {(a) $\lambda$};
			%%%%%%%%%%%%%%%%%%%%%%%%%%%%%%%%%%%%%%%%%%%%
			\begin{scope}[shift={(15,0)}]
				\foreach \x/\y/\z/\w in {0/0/1/0, 0/-1/1/-1, 0/0/0/-1, 1/0/1/-1, 0/-2/1/-2, 0/-1/0/-2, 1/-1/1/-2, 0/-3/1/-3, 0/-2/0/-3, 1/-2/1/-3, 0/-4/1/-4, 0/-3/0/-4, 1/-3/1/-4, 0/-5/1/-5, 0/-4/0/-5, 1/-4/1/-5, 0/-6/1/-6, 0/-5/0/-6, 1/-5/1/-6, 1/0/2/0, 1/-1/2/-1, 2/0/2/-1, 1/-2/2/-2, 2/-1/2/-2, 1/-3/2/-3, 2/-2/2/-3, 1/-4/2/-4, 2/-3/2/-4, 1/-5/2/-5, 2/-4/2/-5,  2/0/3/0, 2/-1/3/-1, 3/0/3/-1, 2/-2/3/-2, 3/-1/3/-2, 2/-3/3/-3, 3/-2/3/-3, 2/-4/3/-4, 3/-3/3/-4, 3/0/4/0, 3/-1/4/-1, 4/0/4/-1, 3/-2/4/-2, 4/-1/4/-2, 3/-3/4/-3, 4/-2/4/-3, 3/-4/4/-4, 4/-3/4/-4, 4/0/5/0, 4/-1/5/-1, 5/0/5/-1, 4/-2/5/-2, 5/-1/5/-2, 4/-3/5/-3, 5/-2/5/-3, 5/0/6/0, 5/-1/6/-1, 6/0/6/-1, 5/-2/6/-2, 6/-1/6/-2, 5/-3/6/-3, 6/-2/6/-3, 6/0/7/0, 6/-1/7/-1, 7/0/7/-1, 6/-2/7/-2, 7/-1/7/-2, 7/-2/6/-3, 7/0/8/0, 7/-1/8/-1, 8/0/8/-1, 8/0/9/0, 8/-1/9/-1, 9/0/9/-1}
				{
					\draw (\x,\y) -- (\z,\w);
				}
				\foreach \x/\y in {0/6, 4/3, 7/1, 9/0}
				{
					\draw[fill=white!40!red] (\x,-\y) -- (\x+1,-\y) -- (\x+1,-\y-1) -- (\x,-\y-1) -- cycle;
				}
				\foreach \x/\y in {1/5, 2/4, 6/2}
				{
					\draw[fill=white!40!red] (\x,-\y) -- (\x+1,-\y) -- (\x+1,-\y-1) -- (\x,-\y-1) -- cycle;
				}
				\node at (4,-8) {(b) ${\rm Add}(\lambda)$};
			\end{scope}
			%%%%%%%%%%%%%%%%%%%%%%%%%%%%%%%%%%%%%%%%%%%%
			\begin{scope}[shift={(30,0)}]
				\foreach \x/\y/\z/\w in {0/0/1/0, 0/-1/1/-1, 0/0/0/-1, 1/0/1/-1, 0/-2/1/-2, 0/-1/0/-2, 1/-1/1/-2, 0/-3/1/-3, 0/-2/0/-3, 1/-2/1/-3, 0/-4/1/-4, 0/-3/0/-4, 1/-3/1/-4, 0/-5/1/-5, 0/-4/0/-5, 1/-4/1/-5, 0/-6/1/-6, 0/-5/0/-6, 1/-5/1/-6, 1/0/2/0, 1/-1/2/-1, 2/0/2/-1, 1/-2/2/-2, 2/-1/2/-2, 1/-3/2/-3, 2/-2/2/-3, 1/-4/2/-4, 2/-3/2/-4, 1/-5/2/-5, 2/-4/2/-5, 2/0/3/0, 2/-1/3/-1, 3/0/3/-1, 2/-2/3/-2, 3/-1/3/-2, 2/-3/3/-3, 3/-2/3/-3, 2/-4/3/-4, 3/-3/3/-4, 3/0/4/0, 3/-1/4/-1, 4/0/4/-1, 3/-2/4/-2, 4/-1/4/-2, 3/-3/4/-3, 4/-2/4/-3, 3/-4/4/-4, 4/-3/4/-4, 4/0/5/0, 4/-1/5/-1, 5/0/5/-1, 4/-2/5/-2, 5/-1/5/-2, 4/-3/5/-3, 5/-2/5/-3, 5/0/6/0, 5/-1/6/-1, 6/0/6/-1, 5/-2/6/-2, 6/-1/6/-2, 5/-3/6/-3, 6/-2/6/-3, 6/0/7/0, 6/-1/7/-1, 7/0/7/-1, 6/-2/7/-2, 7/-1/7/-2, 7/0/8/0, 7/-1/8/-1, 8/0/8/-1, 8/0/9/0, 8/-1/9/-1, 9/0/9/-1}
				{
					\draw (\x,\y) -- (\z,\w);
				}
				\foreach \x/\y in {0/5, 1/4, 5/2, 6/1}
				{
					\draw[fill=\myblue] (\x,-\y) -- (\x+1,-\y) -- (\x+1,-\y-1) -- (\x,-\y-1) -- cycle;
				}
				\foreach \x/\y in {3/3, 8/0}
				{
					\draw[fill=\myblue] (\x,-\y) -- (\x+1,-\y) -- (\x+1,-\y-1) -- (\x,-\y-1) -- cycle;
				}
				\node at (4,-8) {(c) ${\rm Rem}(\lambda)$};
			\end{scope}
		\end{tikzpicture}
		\caption{Sets ${\rm Add}(\star)$ and ${\rm Rem}(\star)$.}
		\label{fig:AddRemYD}
	\end{figure}
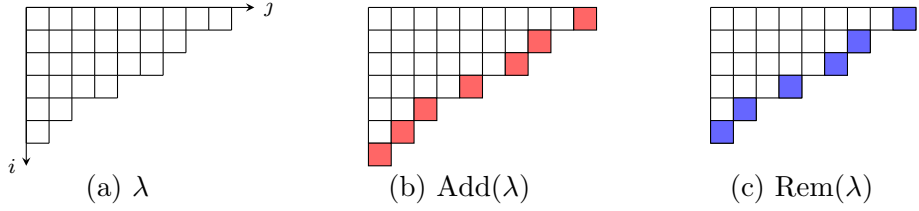
	
	We denote by $\text{Add}(\lambda)$ a subset of border places in diagram $\lambda$ such that one can add a box to that place and the resulting diagram remains a Young diagram. $\text{Rem}(\lambda)$ is the set of boxes in diagram $\lambda$ that one can remove and the resulting diagram remains a Young diagram. An explanatory picture is presented on Fig.\ref{fig:AddRemYD}.

	\subsection{Higher Hamiltonians}
	
	By combining two properties -- the sum over boxes in eigenvalues \eqref{eigenval cut-and-join} and Pierri rules \eqref{Pierri rule 1}, \eqref{Pierri rule 2} -- we could construct a set of nontrivial commuting operators - Hamiltonians for Schur polynomials. 
	At the first step we define the following operators, that generalize the Pierri rule:
	\begin{align}
		\hat{E}_0 &:= p_1 \\
		\hat{E}_1 &:= \Big[ \hat{W}, \hat{E}_0 \Big] = \sum_{a = 1}^{\infty} a \, p_{a + 1} \frac{\p }{\p p_a} \\
		\ldots \\
		\hat{E}_k &:= \Big[ \hat{W}, \hat{E}_{k-1} \Big] 
	\end{align}
	All these operators add one box to the Young diagram yet with different coefficients:
	\begin{equation}
		\label{E adds box}
		\hat{E}_{k} \, S_{\lambda}  = \sum_{\fullbox \in \text{Add}(\lambda)} (\omega_{\fullbox})^k \cdot S_{\lambda + \fullbox}
	\end{equation}
	where we simplify the notation by introducing $\omega_{\Box} := j_{\Box} - i_{\Box}$.
	The above formula could be derived easily step by step in $k$. 
	We demonstrate explicitly only the first step:
	\begin{align}
		\begin{aligned}
			\hat{E}_1 \, S_{\lambda} = \Big[ \hat{W}, \hat{E}_0 \Big] \, S_{\lambda} = \hat{W} \left( \sum_{\fullbox \in \text{Add}(\lambda)} S_{\lambda + \fullbox} \right) - \hat{E}_0 \Big( \kappa_{\lambda} \cdot S_{\lambda} \Big) = \\ =\sum_{\fullbox \in \text{Add}(\lambda)} \kappa_{\lambda + \fullbox} \cdot S_{\lambda + \fullbox} - \kappa_{\lambda} \cdot \sum_{\fullbox \in \text{Add}(\lambda)} S_{\lambda + \fullbox} 
			=\sum_{\fullbox \in \text{Add}(\lambda)} \left( \kappa_{\lambda + \fullbox} - \kappa_{\lambda} \right) \cdot S_{\lambda + \fullbox} = \sum_{\fullbox \in \text{Add}(\lambda)} \omega_{\fullbox} \cdot S_{\lambda + \fullbox} 
		\end{aligned}
	\end{align}
	
	In the same way we define box-removing operators:
	\begin{align}
		\hat{F}_0 &:= \frac{\p }{\p p_1} \\
		\hat{F}_1 &:= - \Big[ \hat{W}, \hat{F}_0 \Big] = \sum_{a = 1}^{\infty} (a+1) \, p_{a} \frac{\p }{\p p_{a+1}} \\
		\ldots \\
		\hat{F}_k &:= - \Big[ \hat{W}, \hat{F}_{k-1} \Big]
	\end{align}
	Operators $F_k$ generalize the second Pierri rule \eqref{Pierri rule 2}:
	\begin{equation}
		\label{F removes box}
		\hat{F}_{k} \, S_{\lambda}  = \sum_{\fullbox \in \text{Rem}(\lambda)} (\omega_{\fullbox})^k \cdot S_{\lambda - \fullbox}
	\end{equation}
	
	On the second step we commute operators $E_k$ and $F_k$ in order to obtain Hamiltonians:
	\begin{equation}
		\hat{H}_{a + b} := \Big[ \hat{E}_a, \hat{F}_b \Big]
	\end{equation}
	There are two nontrivial consequences:
	\begin{itemize}
		\item operators $\Big[ \hat{E}_a, \hat{F}_b \Big]$ depend only on the sum $a + b$ \\
		\item operators $\hat{H}_{k}$ commute
		\begin{equation}
			\Big[ \hat{H}_k, \hat{H}_l \Big] = 0
		\end{equation}
	\end{itemize}
	The above properties follow directly from relations \eqref{E adds box} and \eqref{F removes box}. 
	One could easily prove both properties by showing that Schur polynomials are eigenfunctions of $\Big[ \hat{E}_a, \hat{F}_b \Big]$:
	\begin{equation}
		\label{[e,f]}
		\begin{aligned}
			\Big[ \textcolor{blue}{\hat{E}_a}, \textcolor{red}{\hat{F}_b} \Big] \, S_{\lambda} =  \left( \sum_{\textcolor{red}{\fullbox} \in \text{Rem}(\lambda)} (\omega_{\textcolor{red}{\fullbox}})^{\textcolor{blue}{a} + \textcolor{red}{b}} - \sum_{\textcolor{blue}{\fullbox} \in \text{Add}(\lambda)} (\omega_{\textcolor{blue}{\fullbox}})^{\textcolor{blue}{a} + \textcolor{red}{b}} \right) \cdot S_{\lambda} + \\
			+ \sum_{\substack{\textcolor{red}{\fullbox} \in \text{Rem}(\lambda) \\ \textcolor{blue}{\fullbox} \in \text{Add}(\lambda) \\}} \underbrace{\Big(  (\omega_{\textcolor{blue}{\fullbox}})^{\textcolor{blue}{a}}  (\omega_{\textcolor{red}{\fullbox}})^{\textcolor{red}{b}} S_{\lambda - \textcolor{red}{\fullbox} + \textcolor{blue}{\fullbox}} -  (\omega_{\textcolor{red}{\fullbox}})^{\textcolor{red}{b}} (\omega_{\textcolor{blue}{\fullbox}})^{\textcolor{blue}{a}} S_{\lambda + \textcolor{blue}{\fullbox} - \textcolor{red}{\fullbox}} \Big)}_{0} 
		\end{aligned}
	\end{equation} 
	
	For demonstrative purposes we colored the pieces into \textcolor{blue}{blue} and \textcolor{red}{red} colors, if they correspond to contributions of $\textcolor{blue}{\hat{E}_a}$ and $\textcolor{red}{\hat{F}_a}$ respectively.
	
	The first part of the above relation represents the case when $F_b$ removes box $\Box$ from $\lambda$, and then $E_a$ adds the same (already removed) box $\Box$ to its place thus the resulting diagram remains $\lambda$ as it was in the beginning. The second part is the same as the first part with the difference that $E_a$ add a box to the $\lambda$ and $F_b$ removes it. Therefore, the first and the second parts contribute to the diagonal part of the operator $\Big[ \hat{E}_a, \hat{F}_b \Big]$. 
	
	In contrast, the third and the fourth parts contribute to the off-diagonal part of the operator. In these cases a box is added to the diagram $\lambda$ and the \textit{other} box is removed from the other place of the diagram. In the formula we emphasized $\textcolor{red}{\Box} \in \text{Rem}(\lambda)$ and $\textcolor{blue}{\Box} \in \text{Add}(\lambda)$ meaning $\textcolor{red}{\Box} \not = \textcolor{blue}{\Box}$. In particular, adding and removing different boxes is a commutative operation, therefore:
	\begin{equation}
		\lambda - \textcolor{red}{\Box} + \textcolor{blue}{\Box} = \lambda + \textcolor{blue}{\Box} - \textcolor{red}{\Box},\quad \forall \lambda\neq \varnothing, \Box
	\end{equation}
	
	In general situation cancellation of off-diagonal terms provided by the following condition:
	\begin{equation}
		\label{absent off diagonal}
		\Big[ \hat{E}_0, \hat{F}_0 \Big] = \text{ diagonal operator }
	\end{equation}
	In this particular case $ \Big[ p_1, \frac{\p}{\p p_1} \Big] = -1$. On the other hand we have unit coefficients in Pierri rules \eqref{Pierri rule 1} and \eqref{Pierri rule 2}, this fact directly explains absence of off-diagonal terms.
	
	The above relation explains mutual cancellation of the third and fourth part of \eqref{[e,f]}. 
	Finally, there is the only diagonal part that depends solely on the sum of indices $a + b$, thus Schur polynomials are the common eigenfunctions, and $H_k$ all commute.
	
	The subalgebra of Hamiltonians closes due to the following relations:
	\begin{align}
		\hat{H}_3 &= -6 \cdot \hat{W} \\
		\hat{H}_2 &= -2 \cdot \hat{G}_{\fullbox}
	\end{align}
	that is provided by the peculiar properties of eigenvalues and Young diagrams:
	\begin{align}
		\sum_{\fullbox \in \text{Rem}(\lambda)} \left( \omega_{\fullbox} \right)^3 - \sum_{\fullbox \in \text{Add}(\lambda)} \left( \omega_{\fullbox} \right)^3 &= -6 \cdot \sum_{\fullbox \in \lambda}  \omega_{\fullbox} \\
		\sum_{\fullbox \in \text{Rem}(\lambda)} \left( \omega_{\fullbox} \right)^2 - \sum_{\fullbox \in \text{Add}(\lambda)} \left( \omega_{\fullbox} \right)^2 &= -2 \cdot \sum_{\fullbox \in \lambda} 1
	\end{align}
	that are valid for any Young diagram $\lambda$.
	
	One of the main observations in this paper: the above construction of a commuting system of Hamiltonians can be applied to another family of polynomials that is enumerated by other diagrams. 
	The minimum piece of data that we need:
	\begin{itemize}
		\item an operator that adds boxes to the diagrams (analog of \eqref{Pierri rule 1});
		\item an operator that removes boxes from the diagrams (analog of \eqref{Pierri rule 2});
		\item a Hamiltonian, whose eigenvalues are given by the sum over boxes (analog of \eqref{eigenval cut-and-join}).
	\end{itemize}
	
	\subsection{Additional properties}
	At the end of this section we would like to mention two intriguing properties of Schur Hamiltonians:
	\begin{enumerate}
		\item Operators $\hat{\text{G}}_{\fullbox}$ and $\hat W$ can be considered as the first two non-trivial terms of the following expansion:
		\begin{equation}
			\sum_{n=0}^\infty  S_{[n]}\Big(p_k = kp_k\Big)S_{[n]}\Big( p_k = k^2\frac{\p}{\p p_k}\Big) = 1 + \underbrace{\hat{\text{G}}_{\begin{tikzpicture}[scale=0.15]
						\foreach \i/\j in {0/0, 0/-1}
						{
							\draw[thick] (\i,\j) -- (\i+1,\j);
						}
						\foreach \i/\j in {0/0, 1/0}
						{
							\draw[thick] (\i,\j) -- (\i,\j-1);
						}
						\foreach \i/\j in {}
						{
							\draw[thick] (\i,\j) -- (\i-1,\j-1);
						}
			\end{tikzpicture}}}_{\text{spin} \ 2} + \underbrace{\hat W}_{\text{spin} \ 3} +
			\ldots
			\label{biliSchurs}
		\end{equation}
		where the omitted terms with higher spin (in our notation spin is defined as a total degree in $p_k$ and $\frac{\p}{\p p_k}$ in a normal ordered form), that are {\it no longer} made from the Hamiltonians
		(Schur polynomials are {\it not} their eigenfunctions).
		The powers of $k$ in the arguments of symmetric Schur functions might seem unfamiliar, however they will become apparent after
		a consideration of the Macdonald deformation.
		\item All the constructed Hamiltonians $H_k$ can be expressed in terms of the Schur polynomials themselves and their conjugates with some constant coefficients $A_{\lambda, \mu}$:
		\begin{equation}
			\hat{H}_k = \sum_{|\lambda|= |\mu|} A_{\lambda, \mu} \cdot S_{\lambda}  \, \hat{S}_{\mu}
		\end{equation}
		where we define:
		\begin{equation}
			\hat{S}_{\mu} := S_{\mu}\left( p_k = k \frac{\p}{\p p_k}\right)
		\end{equation}
		There is an \textit{intriguing} property of Hamiltonians $H_k$ that their expansion in terms of the Schur functions include only a sector of single-hook Young diagrams \cite{Mironov:2019mah}.  
		For example, the forms of the grading operator and the cut-and-join operator read:
		\begin{equation}
			\hat{\text{G}}_{\fullbox} = \sum_{a=1}^{\infty} a \, p_a \frac{\partial}{\partial p_a} = \sum_{n=1}^{\infty} \sum_{i,j=0}^{n-1} \ (-)^{i+j} \ S_{[n-i,1^{i}]} \hat{S}_{[n-j,1^{j}]}
		\end{equation}
		\begin{align}
			\begin{aligned}
				\hat W  = -\frac{H_3}{6}=\frac{1}{2}\sum_{a,b=1}^{\infty} (a+b)\cdot p_a p_b \frac{\partial}{\partial p_{a+b}} + a b \cdot p_{a+b}\frac{\partial}{\partial p_{a}} \frac{\partial}{\partial p_{b}}  =\\= \sum_{n=1}^{\infty} \sum_{i,j=0}^{n-1}  \ (-)^{i+j}(n-1-i-j)  \ S_{[n-i,1^{i}]} \hat{S}_{[n-j,1^{j}]}
			\end{aligned}
			\label{Schur Ham Hook expansion}
		\end{align}
		where we denoted a single hook diagram as $[n-i,1^{i}]$, where $i$ is the height of the first column and $n$ is the total number of boxes. 
		The higher $W$-operators could be obtained as proper combinations of $H_k$ \cite{Mironov:2019mah}.
	\end{enumerate}

	\section{Hamiltonians for super-Schur polynomials 
		\label{sSchur}}
	
	Super-diagrams \cite{Galakhov:2023mak, Blondeau-Fournier:2011sft,Noshita:2021dgj} are half-integer partitions $\lambda = \left[ \lambda_1, \lambda_2, \lambda_3, \ldots, \lambda_{l(\lambda)}\right]$, $\lambda_k \in \mathbb{N}/2$ with decreasing rule $\lambda_1 \geqslant \lambda_2 \geqslant \lambda_3 \geqslant \ldots \geqslant \lambda_{l(\lambda)}$. If values $\lambda_k$ and $\lambda_{k+1}$ are both half-integer then inequality necessarily becomes strict $\lambda_k > \lambda_{k+1}$. Usual Young diagrams are the subset of super-Young diagrams. We depict super-partitions as the following diagrams:
	
	\begin{figure}[h!]
		\centering
		\begin{tikzpicture}
			\node(B) at (6,0) {$\begin{array}{c}
					\begin{tikzpicture}[scale=0.5]
						\foreach \i/\j in {0/0, 1/0, 2/0, 3/0, 4/0, 5/0, 6/0, 7/0, 0/-1, 1/-1, 2/-1, 3/-1, 4/-1, 5/-1, 6/-1, 0/-2, 1/-2, 2/-2, 3/-2, 4/-2, 5/-2, 0/-3, 1/-3, 2/-3, 3/-3, 0/-4, 1/-4, 0/-5, 0/-6}
						{
							\draw[thick] (\i,\j) -- (\i+1,\j);
						}
						\foreach \i/\j in {0/0, 0/-1, 0/-2, 0/-3, 0/-4, 0/-5, 1/0, 1/-1, 1/-2, 1/-3, 1/-4, 1/-5, 2/0, 2/-1, 2/-2, 2/-3, 3/0, 3/-1, 3/-2, 4/0, 4/-1, 4/-2, 5/0, 5/-1, 6/0, 6/-1, 7/0 }
						{
							\draw[thick] (\i,\j) -- (\i,\j-1);
						}
						\foreach \i/\j in {8/0, 5/-2, 3/-3}
						{
							\draw[thick] (\i,\j) -- (\i-1,\j-1);
						}
					\end{tikzpicture}
				\end{array}$};
			\node[right] at (B.east) {$=\left[ \frac{15}{2}, 6,\frac{9}{2}, \frac{5}{2}, 1, 1 \right]$};
			\node[left] at (B.west) {$\lambda = $};
		\end{tikzpicture}
		\label{fig:crystalYoung}
	\end{figure}
	The number of super-Young diagrams $p(n,m)$ with $n$ full boxes and $m$ half-boxes can be extracted from the following generating function:
	\begin{equation}
		\prod_{k=1} \frac{1 + y \, x^{k-1}}{1 - x^{k}} = \sum_{n,m=0} p(n,m) \, x^{n} y^{m} = 1 + y + x + 2 x y + 2 x^2 + x y^2 + 4 x^2 y + \ldots
	\end{equation}
	
	If we set $y = 0$ the above generating function becomes generating function for number of usual Young diagrams $p(n)$:
	\begin{equation}
		\prod_{k=1} \frac{1}{1 - x^{k}} = \sum_{n=0} p(n) \, x^{n} = 1 + x + 2 x^2 + 3 x^3 + 5 x^4 + \ldots
	\end{equation}

	\subsection{The simplest Hamiltonians}
	In the case of super-Schur polynomials ordinary time variables $p_a$ are supplemented by Grassmann variables $\theta_a$. Therefore the number of grading operators is increased. 
	There is an operator that counts the number of fermions: 
	\begin{equation}
		\hat{\text{G}}_{\uhbox - \lhbox} = \frac{1}{2}\sum_{k=1}^\infty \theta_k \frac{\p}{\p\theta_k}
	\end{equation}
	with eigenvalues that are given by the sum over half-boxes:
	\begin{equation}
		\label{super grad border}
		\hat{\text{G}}_{\uhbox - \lhbox} \, \mathcal{S}_{\lambda} = \left( \sum_{ \uhbox \in \lambda} \, \frac{1}{2} - \sum_{ \lhbox \in \lambda} \, \frac{1}{2} \right) \cdot  \mathcal{S}_{\lambda}
	\end{equation}
	Note that we treat the full box as a combination of two half-boxes: 
	\begin{equation}
		\begin{tikzpicture}[scale=0.22]
			\foreach \i/\j in {0/0, 0/-1}
			{
				\draw[thick] (\i,\j) -- (\i+1,\j);
			}
			\foreach \i/\j in {0/0,1/0}
			{
				\draw[thick] (\i,\j) -- (\i,\j-1);
			}
			\foreach \i/\j in {}
			{
				\draw[thick] (\i,\j) -- (\i-1,\j-1);
			}
		\end{tikzpicture} = \begin{tikzpicture}[scale=0.22]
			\foreach \i/\j in {0/0, 0/-1}
			{
				\draw[thick] (\i,\j) -- (\i+1,\j);
			}
			\foreach \i/\j in {0/0,1/0}
			{
				\draw[thick] (\i,\j) -- (\i,\j-1);
			}
			\foreach \i/\j in {1/0}
			{
				\draw[thick] (\i,\j) -- (\i-1,\j-1);
			}
		\end{tikzpicture} = \begin{tikzpicture}[scale=0.22]
			\foreach \i/\j in {0/0}
			{
				\draw[thick] (\i,\j) -- (\i+1,\j);
			}
			\foreach \i/\j in {0/0}
			{
				\draw[thick] (\i,\j) -- (\i,\j-1);
			}
			\foreach \i/\j in {1/0}
			{
				\draw[thick] (\i,\j) -- (\i-1,\j-1);
			}
		\end{tikzpicture} + \begin{tikzpicture}[scale=0.22]
			\foreach \i/\j in {0/-1}
			{
				\draw[thick] (\i,\j) -- (\i+1,\j);
			}
			\foreach \i/\j in {1/0}
			{
				\draw[thick] (\i,\j) -- (\i,\j-1);
			}
			\foreach \i/\j in {1/0}
			{
				\draw[thick] (\i,\j) -- (\i-1,\j-1);
			}
		\end{tikzpicture}
	\end{equation}
	Therefore the sum in \eqref{super grad border} effectively runs over non-paired half-boxes $\uhbox$ on the border of the diagram.
	Next grading operator counts half-boxes $\uhbox$ and $\lhbox$ on equal footing:
	
	\begin{equation}
		\hat{\text{G}}_{\uhbox + \lhbox} = \sum_{k=1}^\infty  \left(k \, p_k\frac{\p}{\p p_k} + \Big(k-\frac{1}{2}\Big)\theta_k\frac{\p }{\p \theta_k}\right)
	\end{equation}
	Eigenvalues of this operator have the following form:
	\begin{equation}
		\hat{\text{G}}_{\uhbox + \lhbox} \, \mathcal{S}_{\lambda} = \left(\sum_{ 
			\uhbox \in \lambda} \frac{1}{2} + \sum_{\lhbox \in \lambda} \frac{1}{2} \right) \cdot \mathcal{S}_{\lambda}
	\end{equation}
	From these operators we can construct grading operators that counts only half-boxes of particular kind:
	\begin{equation}
		\hat{\text{G}}_{\uhbox} = \hat{\text{G}}_{\uhbox + \lhbox} + \hat{\text{G}}_{\uhbox - \lhbox} = \sum_{k=1}^\infty  \left(k \, p_k\frac{\p}{\p p_k} + k \, \theta_k\frac{\p }{\p \theta_k}\right)
	\end{equation}
	\begin{equation}
		\hat{\text{G}}_{\lhbox} = \hat{\text{G}}_{\uhbox + \lhbox} - \hat{\text{G}}_{\uhbox - \lhbox} = \sum_{k=1}^\infty  \left(k \, p_k\frac{\p}{\p p_k} + (k-1)\, \theta_k\frac{\p }{\p \theta_k}\right)
	\end{equation}
	where eigenvalues:
	\begin{equation}
		\hat{\text{G}}_{\uhbox} \, \mathcal{S}_{\lambda} = \left(\sum_{
			\uhbox \in \lambda} 1 \right) \cdot \mathcal{S}_{\lambda}
	\end{equation}
	\begin{equation}
		\hat{\text{G}}_{\lhbox} \, \mathcal{S}_{\lambda} = \left(\sum_{
			\lhbox \in \lambda} 1 \right) \cdot \mathcal{S}_{\lambda}
	\end{equation}
	
	Next we provide an explicit form of the first few non-trivial Hamiltonians:
	
	\begin{equation}
		\hat{\cal W}_{\uhbox - \lhbox}= \sum_{a,b=1}^\infty \left(a \theta_{a+b}\frac{\p^2}{\p p_a\p\theta_b} + p_a\theta_b \frac{\p}{\p \theta_{a+b}} \right)
	\end{equation}

	\begin{equation}
		\hat{\cal W}_{\uhbox} = \frac{1}{2}\sum_{a,b=1}^\infty \left(ab p_{a+b} \frac{\p^2}{\p p_a\p p_b} + (a+b)p_ap_b\frac{\p}{\p p_{a+b}}\right)
		+ \sum_{a,b=1}^\infty    b \left(a \theta_{a+b}\frac{\p^2}{\p p_a\p\theta_b} + p_a\theta_b \frac{\p}{\p \theta_{a+b}} \right)
	\end{equation}
	with the eigenvalues
	\begin{equation}
		\hat{\cal W}_{\uhbox-\lhbox} \, \mathcal{S}_{\lambda} = \left(\sum_{ 
			\uhbox \in \lambda} (j_{\uhbox} - i_{\uhbox}) - \sum_{ 
			\lhbox \in \lambda} (j_{\lhbox} - i_{\lhbox})\right) \cdot \mathcal{S}_{\lambda}
	\end{equation}
	
	\begin{equation}
		\hat{\cal W}_{\uhbox} \, \mathcal{S}_{\lambda} = \left(\sum_{ 
			\uhbox \in \lambda} (j_{\uhbox} - i_{\uhbox})  \right) \cdot \mathcal{S}_{\lambda}
	\end{equation}
	A construction of Hamiltonians for only lower-half-boxes is also apparent:
	\begin{equation}
		\hat{\cal W}_{\lhbox} = \hat{\cal W}_{\uhbox} - \hat{\cal W}_{\uhbox - \lhbox}
	\end{equation} 
	\begin{equation}
		\hat{\cal W}_{\lhbox} \, \mathcal{S}_{\lambda} = \left( \sum_{
			\lhbox \in \lambda} (j_{\lhbox} - i_{\lhbox}) \right) \cdot \mathcal{S}_{\lambda}
	\end{equation}
	
	There are similar Pierri rules for super-Schur polynomials. We define four operators
	\begin{align}
		\hat{E}_{0}^{\uhbox} &:= \theta_1 &\hspace{10mm} \hat{E}_{0}^{\lhbox} &:= \sum_{k = 1}^{\infty} p_k \frac{\p}{\p \theta_k} \\
		\hat{F}_{0}^{\uhbox} &:= \frac{\p}{\p \theta_1} &\hspace{10mm} \hat{F}_{0}^{\lhbox} &:= \sum_{k = 1}^{\infty} k \, \theta_k \frac{\p}{\p p_k}
	\end{align}
	that add/remove half-boxes to/from the super-diagrams:
	\begin{align}
		\hat{E}_0^{\shtile} \, \mathcal{S}_{\lambda} &= \sum_{\shtile \in \text{Add}(\lambda)} C_{\lambda,\lambda + \shtile} \, \mathcal{S}_{\lambda + \shtile} &\hspace{5mm} \hat{E}_0^{\shhtile} \, \mathcal{S}_{\lambda} &= \sum_{\shhtile \in \text{Add}(\lambda)} C_{\lambda,\lambda + \shhtile} \, \mathcal{S}_{\lambda + \shhtile} \\
		\hat{F}_0^{\shtile} \, \mathcal{S}_{\lambda} &= \sum_{\shtile \in \text{Rem}(\lambda)} C_{\lambda,\lambda - \shtile} \, \mathcal{S}_{\lambda - \shtile} &\hspace{5mm} \hat{F}_0^{\shhtile} \, \mathcal{S}_{\lambda} &= \sum_{\shhtile \in \text{Rem}(\lambda)} C_{\lambda,\lambda - \shhtile} \, \mathcal{S}_{\lambda - \shhtile} 
	\end{align}
	
	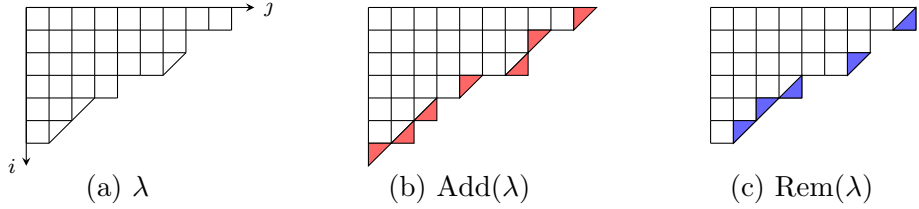
\begin{figure}[ht!]
		\centering
		\begin{tikzpicture}[scale=0.3]
			\foreach \x/\y/\z/\w in {0/0/1/0, 0/-1/1/-1, 0/0/0/-1, 1/0/1/-1, 0/-2/1/-2, 0/-1/0/-2, 1/-1/1/-2, 0/-3/1/-3, 0/-2/0/-3, 1/-2/1/-3, 0/-4/1/-4, 0/-3/0/-4, 1/-3/1/-4, 0/-5/1/-5, 0/-4/0/-5, 1/-4/1/-5, 0/-6/1/-6, 0/-5/0/-6, 1/-5/1/-6, 1/0/2/0, 1/-1/2/-1, 2/0/2/-1, 1/-2/2/-2, 2/-1/2/-2, 1/-3/2/-3, 2/-2/2/-3, 1/-4/2/-4, 2/-3/2/-4, 1/-5/2/-5, 2/-4/2/-5, 2/-5/1/-6, 2/0/3/0, 2/-1/3/-1, 3/0/3/-1, 2/-2/3/-2, 3/-1/3/-2, 2/-3/3/-3, 3/-2/3/-3, 2/-4/3/-4, 3/-3/3/-4, 3/-4/2/-5, 3/0/4/0, 3/-1/4/-1, 4/0/4/-1, 3/-2/4/-2, 4/-1/4/-2, 3/-3/4/-3, 4/-2/4/-3, 3/-4/4/-4, 4/-3/4/-4, 4/0/5/0, 4/-1/5/-1, 5/0/5/-1, 4/-2/5/-2, 5/-1/5/-2, 4/-3/5/-3, 5/-2/5/-3, 5/0/6/0, 5/-1/6/-1, 6/0/6/-1, 5/-2/6/-2, 6/-1/6/-2, 5/-3/6/-3, 6/-2/6/-3, 6/0/7/0, 6/-1/7/-1, 7/0/7/-1, 6/-2/7/-2, 7/-1/7/-2, 7/-2/6/-3, 7/0/8/0, 7/-1/8/-1, 8/0/8/-1, 8/0/9/0, 8/-1/9/-1, 9/0/9/-1}
			{
				\draw (\x,\y) -- (\z,\w);
			}
			\draw[-stealth] (0,0) -- (0,-7);
			\draw[-stealth] (0,0) -- (10,0);
			\node[left] at (0,-7) {$\scriptstyle i$};
			\node[right] at (10,0) {$\scriptstyle j$};
			\node at (4,-8) {(a) $\lambda$};
			%%%%%%%%%%%%%%%%%%%%%%%%%%%%%%%%%%%%%%%%%%%%
			\begin{scope}[shift={(15,0)}]
				\foreach \x/\y/\z/\w in {0/0/1/0, 0/-1/1/-1, 0/0/0/-1, 1/0/1/-1, 0/-2/1/-2, 0/-1/0/-2, 1/-1/1/-2, 0/-3/1/-3, 0/-2/0/-3, 1/-2/1/-3, 0/-4/1/-4, 0/-3/0/-4, 1/-3/1/-4, 0/-5/1/-5, 0/-4/0/-5, 1/-4/1/-5, 0/-6/1/-6, 0/-5/0/-6, 1/-5/1/-6, 1/0/2/0, 1/-1/2/-1, 2/0/2/-1, 1/-2/2/-2, 2/-1/2/-2, 1/-3/2/-3, 2/-2/2/-3, 1/-4/2/-4, 2/-3/2/-4, 1/-5/2/-5, 2/-4/2/-5, 2/-5/1/-6, 2/0/3/0, 2/-1/3/-1, 3/0/3/-1, 2/-2/3/-2, 3/-1/3/-2, 2/-3/3/-3, 3/-2/3/-3, 2/-4/3/-4, 3/-3/3/-4, 3/-4/2/-5, 3/0/4/0, 3/-1/4/-1, 4/0/4/-1, 3/-2/4/-2, 4/-1/4/-2, 3/-3/4/-3, 4/-2/4/-3, 3/-4/4/-4, 4/-3/4/-4, 4/0/5/0, 4/-1/5/-1, 5/0/5/-1, 4/-2/5/-2, 5/-1/5/-2, 4/-3/5/-3, 5/-2/5/-3, 5/0/6/0, 5/-1/6/-1, 6/0/6/-1, 5/-2/6/-2, 6/-1/6/-2, 5/-3/6/-3, 6/-2/6/-3, 6/0/7/0, 6/-1/7/-1, 7/0/7/-1, 6/-2/7/-2, 7/-1/7/-2, 7/-2/6/-3, 7/0/8/0, 7/-1/8/-1, 8/0/8/-1, 8/0/9/0, 8/-1/9/-1, 9/0/9/-1}
				{
					\draw (\x,\y) -- (\z,\w);
				}
				\foreach \x/\y in {0/6, 4/3, 7/1, 9/0}
				{
					\draw[fill=white!40!red] (\x,-\y) -- (\x+1,-\y) -- (\x,-\y-1) -- cycle;
				}
				\foreach \x/\y in {1/5, 2/4, 6/2}
				{
					\draw[fill=white!40!red] (\x+1,-\y) -- (\x+1,-\y-1) -- (\x,-\y-1) -- cycle;
				}
				\node at (4,-8) {(b) ${\rm Add}(\lambda)$};
			\end{scope}
			%%%%%%%%%%%%%%%%%%%%%%%%%%%%%%%%%%%%%%%%%%%%
			\begin{scope}[shift={(30,0)}]
				\foreach \x/\y/\z/\w in {0/0/1/0, 0/-1/1/-1, 0/0/0/-1, 1/0/1/-1, 0/-2/1/-2, 0/-1/0/-2, 1/-1/1/-2, 0/-3/1/-3, 0/-2/0/-3, 1/-2/1/-3, 0/-4/1/-4, 0/-3/0/-4, 1/-3/1/-4, 0/-5/1/-5, 0/-4/0/-5, 1/-4/1/-5, 0/-6/1/-6, 0/-5/0/-6, 1/-5/1/-6, 1/0/2/0, 1/-1/2/-1, 2/0/2/-1, 1/-2/2/-2, 2/-1/2/-2, 1/-3/2/-3, 2/-2/2/-3, 1/-4/2/-4, 2/-3/2/-4, 1/-5/2/-5, 2/-4/2/-5, 2/-5/1/-6, 2/0/3/0, 2/-1/3/-1, 3/0/3/-1, 2/-2/3/-2, 3/-1/3/-2, 2/-3/3/-3, 3/-2/3/-3, 2/-4/3/-4, 3/-3/3/-4, 3/-4/2/-5, 3/0/4/0, 3/-1/4/-1, 4/0/4/-1, 3/-2/4/-2, 4/-1/4/-2, 3/-3/4/-3, 4/-2/4/-3, 3/-4/4/-4, 4/-3/4/-4, 4/0/5/0, 4/-1/5/-1, 5/0/5/-1, 4/-2/5/-2, 5/-1/5/-2, 4/-3/5/-3, 5/-2/5/-3, 5/0/6/0, 5/-1/6/-1, 6/0/6/-1, 5/-2/6/-2, 6/-1/6/-2, 5/-3/6/-3, 6/-2/6/-3, 6/0/7/0, 6/-1/7/-1, 7/0/7/-1, 6/-2/7/-2, 7/-1/7/-2, 7/-2/6/-3, 7/0/8/0, 7/-1/8/-1, 8/0/8/-1, 8/0/9/0, 8/-1/9/-1, 9/0/9/-1}
				{
					\draw (\x,\y) -- (\z,\w);
				}
				\foreach \x/\y in {1/5, 2/4, 6/2}
				{
					\draw[fill=\myblue] (\x,-\y) -- (\x+1,-\y) -- (\x,-\y-1) -- cycle;
				}
				\foreach \x/\y in {3/3, 8/0}
				{
					\draw[fill=\myblue] (\x+1,-\y) -- (\x+1,-\y-1) -- (\x,-\y-1) -- cycle;
				}
				\node at (4,-8) {(c) ${\rm Rem}(\lambda)$};
			\end{scope}
		\end{tikzpicture}
		\caption{Sets ${\rm Add}(\star)$ and ${\rm Rem}(\star)$.}
		\label{fig:AddRem}
	\end{figure}
	
	where $C_{\lambda,\lambda \pm \uhbox}$ and $C_{\lambda,\lambda \pm \lhbox}$ are some constants. The definition of sets $\text{Add}(\lambda)$ and $\text{Rem}(\lambda)$ is clear from the Fig.\ref{fig:AddRem}.
	
	\subsection{Higher Hamiltonians}
	
	To construct higher Hamiltonians for the super-case we follow a procedure similar to the one presented in Sect.\ref{Schur}.
	At the first step we define higher operators recursively:
	\begin{align}
		\hat{E}^{\uhbox}_k &:= \Big[ \hat{\mathcal{W}}_{\uhbox}, \hat{E}^{\uhbox}_{k-1}\Big] &\hspace{10mm} \hat{E}^{\lhbox}_k &:= \Big[ \hat{\mathcal{W}}_{\lhbox}, \hat{E}^{\lhbox}_{k-1}\Big] \\
		\hat{F}^{\uhbox}_k &:= -\Big[ \hat{\mathcal{W}}_{\uhbox}, \hat{F}^{\uhbox}_{k-1}\Big] &\hspace{10mm} \hat{F}^{\lhbox}_k &:= -\Big[ \hat{\mathcal{W}}_{\lhbox}, \hat{F}^{\lhbox}_{k-1}\Big] 
	\end{align} 
	
	At the second step the higher Hamiltonians are defined as follows:
	\begin{equation}
		\hat{H}^{\lhbox}_{a+b} := \Big\{ \hat{E}^{\uhbox}_a, \hat{F}^{\uhbox}_b \Big\} \hspace{20mm} \hat{H}^{\uhbox}_{a+b} := \Big\{ \hat{E}^{\lhbox}_a, \hat{F}^{\lhbox}_b \Big\}
	\end{equation} 
	
	Off-diagonal terms are absent due to relations \eqref{absent off diagonal}:
	\begin{equation}
		\Big\{ \hat{E}_{0}^{\uhbox}, \hat{F}_{0}^{\uhbox} \Big\} = 1
	\end{equation}
	
	\begin{equation}
		\Big\{ \hat{E}_{0}^{\lhbox}, \hat{F}_{0}^{\lhbox} \Big\} = \sum_{k=1}^{\infty} \left( k \, p_k \frac{\p}{\p p_k} + k \, \theta_k \frac{\p}{\p \theta_k} \right)
	\end{equation}
	
	Here anti-commutators are implemented rather than the usual commutators due to Grassmann odd nature of $\hat{E}^{\uhbox, \lhbox}_k$ and $\hat{F}^{\uhbox, \lhbox}_k$ operators.
	Note that the labels of operators in the r.h.s. are different from the labels of Hamiltonians in the l.h.s. 
	These Hamiltonians form a commuting set:
	\begin{align}
		\Big[ \hat{H}^{\uhbox}_{k}, \hat{H}^{\uhbox}_{l} \Big] = 0 \\
		\Big[ \hat{H}^{\lhbox}_{k}, \hat{H}^{\lhbox}_{l} \Big] = 0 \\
		\Big[ \hat{H}^{\uhbox}_{k}, \hat{H}^{\lhbox}_{l} \Big] = 0
	\end{align}
	
	This a bit awkwardly looking notation for Hamiltonians is determined by the following relations that make the subalgebra closed:
	\begin{align}
		\hat{H}_2^{\lhbox} = \hat{\text{G}_{\lhbox}} \hspace{20mm} \hat{H}_3^{\lhbox} = 2 \cdot \hat{\mathcal{W}_{\lhbox}} \\
		\hat{H}_0^{\uhbox} = \hat{\text{G}_{\uhbox}} \hspace{20mm} \hat{H}_1^{\uhbox} = 2 \cdot \hat{\mathcal{W}_{\uhbox}} 
	\end{align}
	\subsection{Additional properties}
	\begin{enumerate}
		\item

		Since  super-Schurs for integer diagrams (not containing unpaired upper-half-boxes) coincide with the ordinary Schurs  (do not depend on $\theta_k$) from the following combinations we obtain usual grading operator and cut-and-join operator: 
		\begin{equation}
			\sum_{n=0}^\infty  {\cal S}_{[n]}\Big(p_k=  kp_k\Big){\cal S}_{[n]}\Big(p_k= k^2\frac{\p}{\p p_k}\Big) = 1 + \hat{\text{G}}_{\fullbox} + \hat W +
			\ldots
		\end{equation}

		However, a similar combination for half-integer diagrams gives the following result:
		\begin{equation}
			\begin{aligned}
				\sum_{n\ {\rm odd}} {\cal S}_{[n/2]}\Big(p_k = kp_k, \theta_k = k\theta_k\Big)
				{\cal S}_{[n/2]}\Big( p_k = k^2\frac{\p}{\p p_k}, \theta_k = k\frac{\p}{\p\theta_k}\Big) = \\
				= \sum_{k=1}^\infty k\theta_{k}\frac{\p}{\p\theta_k}
				+  \sum_{a,b=1}^\infty    b \left(a \theta_{a+b}\frac{\p^2}{\p p_a\p\theta_b} + p_a\theta_b \frac{\p}{\p \theta_{a+b}} \right)
				+ \ldots
			\end{aligned}
			\label{semiintegersupershursum}
		\end{equation}
		(note that $\frac{\p}{\p \theta_k}$ in the argument is multiplied by $k$, not by $k^2$).
		Operators as being split in brackets in the r.h.s. are {\it not} grading operators, and super-Schurs are {\it not}  eigenfunctions
		of each of them separately.
		Yet the r.h.s. \emph{is} a combination of the Hamiltonians:
		\begin{equation}
			\sum_{n=0}^\infty {\cal S}_{[n/2]}\Big(p_k = kp_k, \theta_k = k\theta_k\Big)
			{\cal S}_{[n/2]}\Big( p_k = k^2\frac{\p}{\p p_k}, \theta_k = k\frac{\p}{\p\theta_k}\Big)
			= 1 + \hat{\text{G}}_{\uhbox} + \hat{\cal W}_{\uhbox} + \ldots
		\end{equation}
		
		Operators $\hat{\text{G}}_{\lhbox}$ and $\hat{\mathcal{W}}_{\lhbox}$ in bilinear combination of super-Schurs remain to be found.
		
		\item If we expand this operators in terms of super-Schur polynomials and their conjugates:
		\begin{equation}
			\hat{\mathcal{S}}_{\lambda} := \mathcal{S}_{\lambda} \left( p_k = k \frac{\p }{\p p_k}, \theta_k = \frac{\p }{\p \theta_k}\right)
		\end{equation}
		we observe the same \textit{miraculous} property as in \eqref{Schur Ham Hook expansion}: only single-hook diagrams contribute. 
		
		However, in this case there are some peculiarities. 
		We do not know yet any natural normalization conditions for the super-Schur polynomials. 
		We choose the normalization with unit coefficients in front of monomials $p_1^n$ and $p_1^n \theta_1$ respectively. Therefore, the orthogonality conditions have the following form:
		\begin{equation}
			\hat{\mathcal{S}}_{\lambda} \, \mathcal{S}_{\mu} = \delta_{\lambda, \mu} \cdot ||\mathcal{S}_{\lambda}||^2
		\end{equation}
		with some coefficients $||\mathcal{S}_{\lambda}||^2$.
		
		Moreover, in case of super-diagrams we have four types of single-hook diagrams. 
		Two types on even levels $|\lambda| = 2m$ (B and F stand for boson and fermion respectively):
		\begin{equation}
			\text{FF type}: \left[\frac{k}{2}, 1^l, \frac{1}{2} \right] \hspace{20mm} \text{BB type}: [k, 1^l]
		\end{equation}
		For example, $|\lambda| = 5$:
		\begin{align}
			\left[\frac{7}{2}, 1, \frac{1}{2} \right] = \begin{array}{c}
				\begin{tikzpicture}[scale=0.3]
					\foreach \i/\j in {0/0, 0/-1, 0/-2, 1/0, 1/-1, 2/0, 2/-1, 3/0}
					{
						\draw[thick] (\i,\j) -- (\i+1,\j);
					}
					\foreach \i/\j in {0/0, 1/0, 0/-1, 1/-1, 2/0, 0/-2, 3/0}
					{
						\draw[thick] (\i,\j) -- (\i,\j-1);
					}
					\foreach \i/\j in {1/-2, 4/0}
					{
						\draw[thick] (\i,\j) -- (\i-1,\j-1);
					}
				\end{tikzpicture}
			\end{array}
			\hspace{20mm}
			[3,1^2] = \begin{array}{c}
				\begin{tikzpicture}[scale=0.3]
					\foreach \i/\j in {0/0, 0/-1, 0/-2, 1/0, 1/-1, 2/0, 0/-3, 2/-1}
					{
						\draw[thick] (\i,\j) -- (\i+1,\j);
					}
					\foreach \i/\j in {0/0, 1/0, 0/-1, 1/-1, 2/0, 0/-2, 1/-2, 3/0}
					{
						\draw[thick] (\i,\j) -- (\i,\j-1);
					}
					\foreach \i/\j in {}
					{
						\draw[thick] (\i,\j) -- (\i-1,\j-1);
					}
				\end{tikzpicture}
			\end{array}
		\end{align}
		And two types on odd levels $|\lambda| = 2m+1$:
		\begin{equation}
			\text{FB type}:\left[\frac{k}{2}, 1^l\right] \hspace{20mm} \text{BF type}:\left[k, 1^l,\frac{1}{2}\right]
		\end{equation}
		For example, $|\lambda| = \frac{9}{2}$:
		\begin{align}
			\left[\frac{5}{2}, 1^2\right] = \begin{array}{c}
				\begin{tikzpicture}[scale=0.3]
					\foreach \i/\j in {0/0, 0/-1, 0/-2, 1/0, 1/-1, 2/0, 0/-3}
					{
						\draw[thick] (\i,\j) -- (\i+1,\j);
					}
					\foreach \i/\j in {0/0, 1/0, 0/-1, 1/-1, 2/0, 0/-2, 1/-2}
					{
						\draw[thick] (\i,\j) -- (\i,\j-1);
					}
					\foreach \i/\j in {3/0}
					{
						\draw[thick] (\i,\j) -- (\i-1,\j-1);
					}
				\end{tikzpicture}
			\end{array} 
			\hspace{20mm}
			\left[2, 1^2,\frac{1}{2}\right] = \begin{array}{c}
				\begin{tikzpicture}[scale=0.3]
					\foreach \i/\j in {0/0, 0/-1, 0/-2, 1/0, 1/-1, 0/-3}
					{
						\draw[thick] (\i,\j) -- (\i+1,\j);
					}
					\foreach \i/\j in {0/0, 1/0, 0/-1, 1/-1, 2/0, 0/-2, 1/-2, 0/-3}
					{
						\draw[thick] (\i,\j) -- (\i,\j-1);
					}
					\foreach \i/\j in {1/-3}
					{
						\draw[thick] (\i,\j) -- (\i-1,\j-1);
					}
				\end{tikzpicture}
			\end{array}
		\end{align}
		Finally, the following single-hook expansion holds for super-Schur Hamiltonians:
		\begin{equation}
			\hat{\cal W}_{\uhbox} = \hat{\cal W}^{(1)}_{\uhbox} + \hat{\cal W}^{(2)}_{\uhbox} +\hat{\cal W}^{(3)}_{\uhbox}+ \hat{\cal W}^{(4)}_{\uhbox}+ \hat{\cal W}^{(5)}_{\uhbox}
		\end{equation}
		\begin{align}
			\label{single-hook expansion sSchurs Ham}
			\begin{aligned}
				\text{BB} \cdot \hat{\text{BB}}:& \\
				&\hat{\cal W}^{(1)}_{\uhbox} = \sum_{n \ \text{even} } \ \sum_{i,j = 0}^{\frac{n}{2}-1} (-)^{i+j} \left( \frac{n-2}{2} - i - j \right) \cdot \frac{C^{j}_{\frac{n-1}{2}}}{C^{i}_{\frac{n-1}{2}}} \cdot \frac{\mathcal{S}_{[\frac{n}{2} -i, 1^{i}]} \, \hat{\mathcal{S}}_{[\frac{n}{2}-j, 1^{j}]}}{|| \mathcal{S}_{[\frac{n}{2} -i, 1^{i}]}||^2} \  \\
				\text{FB} \cdot \hat{\text{FB}}:& \\
				&\hat{\cal W}^{(2)}_{\uhbox} = \sum_{n \ \text{odd} } \ \sum_{i,j = 0}^{\frac{n-3}{2}} (-)^{i+j} \left( \frac{n-1}{2} - i - j \right) \cdot \frac{C^{j}_{\frac{n-3}{2}}}{C^{i}_{\frac{n-1}{2}}} \cdot \frac{\mathcal{S}_{[\frac{n}{2} -i , 1^{i}]} \, \hat{\mathcal{S}}_{[\frac{n}{2}-j, 1^{j}]}}{|| \mathcal{S}_{[\frac{n}{2} -i, 1^{i}]}||^2} \  \\
				\text{BF} \cdot \hat{\text{BF}}:& \\
				&\hat{\cal W}^{(3)}_{\uhbox} = \sum_{n \ \text{odd} } \ \sum_{i,j = 0}^{\frac{n-3}{2}} (-)^{i+j} \left( \frac{n-5}{2} - i - j \right) \cdot \frac{C^{j}_{\frac{n-3}{2}}}{C^{i+1}_{\frac{n-1}{2}}} \cdot \frac{ \mathcal{S}_{[\frac{n-1}{2}-i, 1^{i},\frac{1}{2}]} \, \hat{\mathcal{S}}_{[\frac{n-1}{2}-j, 1^{j},\frac{1}{2}]}}{|| \mathcal{S}_{[\frac{n-1}{2}-i, 1^{i},\frac{1}{2}]}||^2} \  \\
				\text{FB} \cdot \hat{\text{BF}}:& \\ 
				&\hat{\cal W}^{(4)}_{\uhbox} = \sum_{n \ \text{odd} } \ \sum_{i,j = 0}^{\frac{n-3}{2}} (-)^{i+j+1} \left( \frac{n-3}{2} - i - j \right) \cdot \frac{C^{j}_{\frac{n-3}{2}}}{C^{i}_{\frac{n-1}{2}}} \cdot \frac{\mathcal{S}_{[\frac{n}{2} -i, 1^{i}]} \, \hat{\mathcal{S}}_{[\frac{n-1}{2}-j, 1^{j},\frac{1}{2}]}}{|| \mathcal{S}_{[\frac{n}{2} -i, 1^{i}]}||^2} \  \\
				\text{BF} \cdot \hat{\text{FB}}:& \\ 
				&\hat{\cal W}^{(5)}_{\uhbox} = \sum_{n \ \text{odd} } \ \sum_{i,j = 0}^{\frac{n-3}{2}} (-)^{i+j+1} \left( \frac{n-3}{2} - i - j \right) \cdot \frac{C^{j}_{\frac{n-3}{2}}}{C^{i+1}_{\frac{n-1}{2}}} \cdot \frac{ \mathcal{S}_{[\frac{n-1}{2}-i, 1^{i},\frac{1}{2}]} \, \hat{\mathcal{S}}_{[\frac{n}{2}-j, 1^{j}]} }{|| \mathcal{S}_{[\frac{n-1}{2}-i, 1^{i},\frac{1}{2}]}||^2} \ 
			\end{aligned}
		\end{align} 
		Note that for this Hamiltonian sector $\text{FF}$ never enters. The other super-Hamiltonian $\hat{\mathcal{W}_{\lhbox}}$ for super-Schurs also have single-hook expansion, although we do not present it here. In general, higher super-Hamiltonians $\hat{H}_k^{\uhbox, \lhbox}$ do not have neither single-hook expansion nor "sum over boxes" property for eigenvalues. However, special combinations of $\hat{H}_k^{\uhbox, \lhbox}$ may have these properties, we postpone this question for future research.
	\end{enumerate}

	\section{Hamiltonians for Macdonald polynomials
		\label{Macs}}
	
	\subsection{Differential grading operator and finite-difference/shift cut-and-join}
	
	Now in addition to the grading operator appeared in the story of the Schur polynomials in Sec.~\ref{Schur}, which remains a differential operator,
	\begin{equation}
		\hat{\text{G}}_{\fullbox} = \sum_{k=1}^\infty k \, p_k\frac{\p}{\p p_k}
	\end{equation}
	\begin{equation}
		\hat{\text{G}}_{\fullbox} \, M^{q,t}_{\lambda} = \left( \sum_{\fullbox \in \lambda} 1 \right) \cdot M^{q,t}_{\lambda}
	\end{equation}
	we have non-local (shift) Hamiltonians \cite{Mironov:2019uoy}, the first and the simplest one has the following vertex operator form 
	\begin{equation}
		\label{Macdonald H}
		\hat{  \mathcal{H}}^{+}_{\fullbox} = \oint \frac{dz}{z} \exp \left( \sum^{\infty}_{k = 1} \frac{(1-t^{-2k})}{k} \, p_k \, z^k \right) \,
		\exp \left(\sum^{\infty}_{k = 1} \frac{(q^{2k} - 1)}{z^k} \, \frac{\partial}{\partial p_k} \right)
	\end{equation}
	
	This single Hamiltonian turns out to be sufficient to determine all the Macdonald polynomials $M^{q,t}_{\lambda}$ as eigenfunctions with distinct eigenvalues:
	\begin{align}
		\hat{  \mathcal{H}}^{+}_{\fullbox} \, M^{q,t}_{\lambda} = \mathcal{E}^{\, q,t}_{\lambda} \, M^{q,t}_{\lambda}
	\end{align}
	An expression for eigenvalue $\mathcal{E}^{\, q,t}_{\lambda}$ has a form of a sum over all boxes in the Young diagram:
	\begin{equation}
		\mathcal{E}^{\, q,t}_{\lambda} = 1 + (q^2 - 1)(1 - t^{-2}) \cdot \sum_{\Box \in \lambda} q^{2 j_{\Box}} t^{-2 i_{\Box}}
	\end{equation}
	where the coordinate of the initial box in the Young diagram reads $(i_{\Box}, j_{\Box}) = (0,0)$ and $i_{\Box}$, $j_{\Box}$ are vertical and horizontal coordinates correspondingly. 
	
	\subsection{Single-hook Schur decomposition of Macdonald Hamiltonian}
	
	With the help of reduced Cauchy formula
	\begin{equation}
		\sum_{n=0}^\infty S_{[n]}\{p\}\cdot z^n = \exp\left(\sum_k \frac{p_kz^k}{k}\right)
	\end{equation}
	this Hamiltonian can be rewritten as a much healthier analogue of (\ref{biliSchurs}):
	\begin{equation} \label{MHamdeco}
		\hat{  \mathcal{H}}^{+}_{\fullbox} = \sum_{n = 0}^\infty S_{[n]} \Big(  p_k = (1 - t^{-2k})\cdot p_k \Big) \cdot S_{[n]} \Big( p_k = k \cdot (q^{2k} - 1)\cdot \frac{\partial}{\partial p_k} \Big)
	\end{equation}
	where we can further substitute the single-hook expansion \cite{Mironov:2019mah}
	\begin{equation}
		S_{[n]} \Big(  p_k = (1 - t^{-2k})\cdot p_k \Big) = (1-t^{-2})
		\sum_{i=0}^{n-1} (-)^i\cdot t^{-2i}\cdot S_{[n-i,1^i]}\{p_k\}
	\end{equation}
	to arrive to the following expression:
	\begin{equation}
		\hat{  \mathcal{H}}^{+}_{\fullbox} = 1 +(q^2 - 1)(1-t^{-2})\ \sum_{n=1}^\infty \ \sum_{i,j=0}^{n-1} (-)^{i+j}\cdot t^{-2i}q^{2(n-1-j)}\cdot
		S_{[n-i,1^i]} \cdot \hat{S}_{[n-j,1^j]}
		\label{MHamdeco1}
	\end{equation}
	Note that symmetric and single-hook decompositions \eqref{MHamdeco} and \eqref{MHamdeco1} are exact,
	all terms in the sums should be taken into account -- not like it was in the case of \eqref{biliSchurs}.
	
	\subsection{$q,t$-inversion symmetry}
	
	The Macdonald polynomials have symmetry with respect to an inversion of $q$ and $t$ parameters: $M_\lambda^{q,t} = M_\lambda^{q^{-1},t^{-1}}$.
	Even more surprisingly, this property is not reflected by the Hamiltonian (\ref{Macdonald H}),
	which implies there should be another Hamiltonian obtained by this transformation with the same eigenfunctions:
	\begin{equation}
		\hat{  \mathcal{H}}^{-}_{\fullbox} = \oint \frac{dz}{z} \exp \left( \sum_{k = 1} \frac{(1-t^{2k})}{k} \, p_k \, z^k \right) \,
		\exp \left(\sum_{k = 1} \frac{(q^{-2k} - 1)}{z^k} \, \frac{\partial}{\partial p_k} \right)
	\end{equation}
	\begin{equation}
		\hat{  \mathcal{H}}^{-}_{\fullbox} M_\lambda^{q,t} = \tilde{\cal E}_\lambda^{q,t} M_\lambda^{q,t}
	\end{equation}
	\begin{equation}
		\tilde{\cal E}_\lambda^{q,t} =  {\cal E}_\lambda^{\,q^{-1} \!\! ,\,   t^{-1}}
		= 1 + (q^{-2} - 1)(1 - t^{2}) \cdot \sum_{\Box \in \lambda} q^{-2 j_{\Box}} t^{2 i_{\Box}},
	\end{equation}
	\subsection{Higher Hamiltonians}
	Operators $p_1$ and $\frac{\p }{\p p_1}$ add and remove boxes in Macdonald polynomials:
	\begin{equation}
		p_1 \cdot M^{q,t}_{\lambda} = \sum_{\fullbox \in \text{Add}(\lambda)} C^{q,t}_{\lambda, \lambda + \fullbox} \ M^{q,t}_{\lambda + \fullbox}
	\end{equation}
	\begin{equation}
		\frac{\p}{\p p_1} \ M^{q,t}_{\lambda} = \sum_{\fullbox \in \text{Rem}(\lambda)} C^{q,t}_{\lambda, \lambda - \fullbox} \ M^{q,t}_{\lambda - \fullbox}
	\end{equation}
	where coefficients $C_{\lambda, \lambda + \Box}^{q,t}$ and $C_{\lambda, \lambda - \Box}^{q,t}$ are rational functions of parameters $q,t$.
	With the help of these two Hamiltonians one can define operators $\hat{E}_k$, $\hat{F}_k$ recursively for all integer values $k \in \mathbb{Z}$:
	\begin{equation}
		\hat{E}_{k+1} := \Big[ \hat{\mathcal{H}}^{+}_{\fullbox}, \hat{E}_k \Big]  \hspace{10mm} \hat{F}_{k+1} := -\Big[ \hat{\mathcal{H}}^{+}_{\fullbox}, \hat{F}_k \Big] 
	\end{equation}
	\begin{equation}
		\hat{E}_{k-1} := \Big[ \hat{\mathcal{H}}^{-}_{\fullbox}, \hat{E}_k \Big]  \hspace{10mm} \hat{F}_{k-1} := -\Big[ \hat{\mathcal{H}}^{-}_{\fullbox}, \hat{F}_k \Big] 
	\end{equation}
	Initial conditions for these recursion $\hat{E}_{0} = p_1$ and $\hat{F}_{0} = \frac{\p}{\p p_1}$.
	Action of these operators on Macdonald polynomials reads:
	\begin{equation}
		\hat{E}_{k} \ M^{q,t}_{\lambda} = \sum_{\fullbox \in \text{Add}(\lambda)} \ \left( q^{2j_{\fullbox}} t^{-2i_{\fullbox}}\right)^{k} \ C^{q,t}_{\lambda, \lambda + \fullbox} \ M^{q,t}_{\lambda + \fullbox}
	\end{equation}
	\begin{equation}
		\hat{F}_{k} \ M^{q,t}_{\lambda} = \sum_{\fullbox \in \text{Rem}(\lambda)} \ \left( q^{2j_{\fullbox}} t^{-2i_{\fullbox}}\right)^{k} \ C^{q,t}_{\lambda, \lambda - \fullbox} \ M^{q,t}_{\lambda - \fullbox}
	\end{equation}
	Then Hamiltonians reads:
	\begin{equation}
		\hat{\mathcal{H}}^{\prime}_{a+b} := \Big[ \hat{E}_{a}, \hat{F}_{b} \Big] 
	\end{equation}
	that form commuting family:
	\begin{equation}
		\Big[ \hat{\mathcal{H}}^{\prime}_{a}, \hat{\mathcal{H}}^{\prime}_{b} \Big] = 0
	\end{equation}
	Higher Hamiltonians from \cite{Mironov:2019uoy} should be expressed as proper combinations of our $\hat{\mathcal{H}}^{\prime}_{a}$. We leave it question for future papers.

	\section{Hamiltonians for super-Macdonald polynomials 
		\label{sMac}}
	
	Super-Macdonald polynomials $\mathcal{M}_{\lambda}^{q,t}$ \cite{GMT2, GMT3, Blondeau-Fournier:2011sft, Blondeau-Fournier:2012exj} are $q,t$-deformations of super-Schur polynomials therefore the two grading operators are just the same:
	\begin{equation}
		\hat{\text{G}}_{\uhbox} \, \mathcal{M}^{q,t}_{\lambda} = \left(\sum_{
			\uhbox \in \lambda} 1 \right) \cdot \mathcal{M}_{\lambda}^{q,t}
	\end{equation}
	\begin{equation}
		\hat{\text{G}}_{\lhbox} \, \mathcal{M}_{\lambda}^{q,t} = \left(\sum_{
			\lhbox \in \lambda} 1 \right) \cdot \mathcal{M}_{\lambda}^{q,t}
	\end{equation}
	
	In complete analogy with the case of canonical Macdonald polynomials where the Hamiltonian \eqref{Macdonald H} have eigenvalues:
	\begin{equation}
		\mathcal{E}^{\, q,t}_{\lambda} = 1 + (q^2 - 1)(1 - t^{-2}) \cdot \sum_{\Box \in \lambda} q^{2 j_{\Box}} t^{-2 i_{\Box}}
	\end{equation}
	we define new super-Hamiltonians for super-Macdonald polynomials \cite{GMT2, GMT3, Blondeau-Fournier:2011sft, Blondeau-Fournier:2012exj}:
	\begin{equation}
		\label{def sup eigval}
		\begin{aligned}
			&\hat{\mathcal{H}}^{\shtile,+}  \mathcal{M}^{q,t}_{\lambda}=\left[1+(q^2-1)(1-t^{-2})\left(\sum\limits_{\shtile\in\lambda}q^{2j_{\shtile}}t^{-2i_{\shtile}}\right)\right] \mathcal{M}^{q,t}_{\lambda} \\
			&\hat{\mathcal{H}}^{\shhtile,+}  \mathcal{M}^{q,t}_{\lambda}=\left[1+(q^2-1)(1-t^{-2})\left(\sum\limits_{\shhtile\in\lambda} q^{2j_{\shhtile}} t^{-2i_{\shhtile}}\right)\right]\mathcal{M}^{q,t}_{\lambda}\\
			&\hat{\mathcal{H}}^{\shtile,-}  \mathcal{M}^{q,t}_{\lambda}=\left[1+(q^{-2}-1)(1-t^{2})\left(\sum\limits_{\shtile\in\lambda}q^{-2j_{\shtile}}t^{2i_{\shtile}}\right)\right]\mathcal{M}^{q,t}_{\lambda}\\
			&\hat{\mathcal{H}}^{\shhtile,-}  \mathcal{M}^{q,t}_{\lambda}=\left[1+(q^{-2}-1)(1-t^{2})\left(\sum\limits_{\shhtile\in\lambda}t^{2x_{\shhtile}}q^{-2y_{\shhtile}}\right)\right]\mathcal{M}^{q,t}_{\lambda}\\
		\end{aligned}
	\end{equation}
	For even (ordinary) diagrams the eigenvalues of the first two Hamiltonians coincide, whereas those of the last two
	are obtained by a switch $q \to q^{-1}, t \to t^{-1}$. 
	However, for a generic odd diagram all the four eigenvalues are different and independent. 
	
	Based on super-Macdonald polynomials \cite{GMT2, GMT3, Blondeau-Fournier:2011sft, Blondeau-Fournier:2012exj} and eigenvalues \eqref{def sup eigval} we compute explicit form of the super-Hamiltonians and propose the following conjectural formulas \eqref{sHam1} and \eqref{sHam2} for super-Hamiltonians. We have checked our formulas for super-Hamiltonians up to diagrams of order $|\lambda| = \frac{25}{2}$.
	
	For the first pair of Hamiltonians we present the following closed expression:
	\begin{tcolorbox}
		\begin{equation}
			\label{sHam1}
			\begin{aligned}
				&\hat{\mathcal{H}}^{\shtile,+}= \oint \frac{dw}{w} \bra{\varnothing}  \left[\exp\left\{\sum\limits_{k=1}^{\infty}\left(\frac{1-t^{-2k}}{k}p_kw^{2k}
				+(1-t^{-2}) {t^{-2(k-1)}}w^{2k-1}\theta_{k}\sum\limits_{m=0}^{k-1}t^{2m}\psi_m\right)\right\} \times 
				\right.\\ 
				&\left. \times\exp\left\{\sum\limits_{k=1}^{\infty}\left((q^{2k}-1)\frac{\p}{\p p_k}w^{-2k}+(q^2-1) {q^{2(k-1)}}w^{-2k+1}\,\frac{\p}{\p \theta_{k}}\sum\limits_{m=0}^{k-1}q^{-2m}\psi_m^{\dagger}\right)\right\}\right]\ket{\varnothing}_F\,,\\
				&\hat{\mathcal{H}}^{\shhtile,+}=\oint \frac{dw}{w} \bra{\varnothing} \left[\exp\left\{\sum\limits_{k=1}^{\infty} \frac{1-t^{-2k}}{k}p_kw^{2k}+
				\sum\limits_{ {k=2}}^{\infty}\left((1-t^{-2})
				{t^{-2(k-1)}}  w^{2k-1}\theta_{k}\sum\limits_{m=0}^{k-2}t^{2m}\psi_m\right) \right\}\right.\\
				&\left.\times\exp\left\{\sum\limits_{k=1}^{\infty}(q^{2k}-1)\frac{\p}{\p p_k}w^{-2k}
				+\sum\limits_{ {k=2}}^{\infty}\left((q^2-1) q^{2(k-2)}w^{-2k+1}\,\frac{\p}{\p \theta_{k}}\sum\limits_{m=0}^{k-2}q^{-2m}\psi_m^{\dagger}\right)\right\}\right]\ket{\varnothing}_F\,,\\
			\end{aligned}
		\end{equation}
	\end{tcolorbox}
	where fermionic correlator are defined by:
	\begin{equation}
		\left\{\psi_i,\psi_j^{\dagger}\right\}=\delta_{ij},\quad \psi_i\ket{\varnothing}_F=0,\quad \langle\varnothing|\varnothing\rangle_F=1
	\end{equation}
	For the second pair of Hamiltonians we propose the following expressions:
	\begin{tcolorbox}
		\begin{equation}
			\label{sHam2}
			\begin{aligned}
				&\hat{\mathcal{H}}^{\shtile,-}=\oint \frac{dw}{w} \bra{\varnothing} \left[
				\exp\left\{\sum\limits_{k=1}^{\infty}\left(\frac{1-t^{2k}}{k}p_kw^{2k}s^{-k}
				+(1-t^2)w^{2k-1}\,\theta_{k} \, \nu \right)\right\}\times\right.\\
				&\left.\times\exp\left\{\sum\limits_{k=1}^{\infty}\left((q^{-2k}-1)\frac{\p}{\p p_k}w^{-2k}+(q^{-2}-1)w^{-2k+1}\,\frac{\p}{\p \theta_{k}} \nu^\dagger s^{k-1}\right)\right\}\right]\ket{\varnothing}_B\,,\\
				&\hat{\mathcal{H}}^{\shhtile,-}=\oint \frac{dw}{w} \bra{\varnothing} \left[
				\exp\left\{\sum\limits_{k=1}^{\infty}\left(\frac{1-t^{2k}}{k}p_kw^{2k}s^{-k}
				+(1-t^2)w^{2k-1}\,\theta_{k} \, \nu \right)\right\}\times\right.\\
				&\left.\times\exp\left\{\sum\limits_{k=1}^{\infty}\left((q^{-2k}-1)\frac{\p}{\p p_k}w^{-2k}
				+(q^{-2}-1)w^{-2k+1}\,\frac{\p}{\p \theta_{k}} \, \nu^\dagger s^{k-2}\right)\right\} \right]\Big|\ket{\varnothing}_B
			\end{aligned}
		\end{equation}
	\end{tcolorbox}

	where $s$ is an auxiliary invertible bosonic operator, whereas $\nu,\nu^\dagger$ are anti-commutative fermion operators, that commute $[ s, \nu] = [s, \nu^{\dagger}] = 0$. 
	For their correlators we define:
	\begin{equation}
		\bra{\varnothing} s^a \ket{\varnothing}_B=1,\quad \bra{\varnothing} \nu \nu^{\dagger} s^b \ket{\varnothing}_B=\left\{\begin{array}{ll}
			\dfrac{1-(qt)^{2b+2}}{q^{2b}(1-(qt)^{2})}, & \mbox{if }b\geq 0\,;\\
			0,& \mbox{otherwise}
		\end{array}\right.
	\end{equation}
    Here we use the correlator description to construct combinatorially a generic term of an infinite series. Afterwards the infinite series expression could be "shrunk" into compact expression \eqref{sHam2}. It would be natural to wonder if newly introduced fields $s$ and $\nu$ have a deeper physical meaning, as in \eqref{sHam1} $z$ and $\psi$ were "dual" to ordinary and Grassman time variables. However, unfortunately, at the moment we are unaware of such meaning and use them as a mere simplification trick. Note that there is a single fermionic pair $\nu,\nu^\dagger$, therefore these two Hamiltonians contain only terms 
	with at most the first power of $\theta$ and $\theta$-derivative.
	The primary role of $s$ is to exclude terms like $p_{k_1}\ldots p_{k_n}\frac{\p}{\p\theta_{m-\frac{1}{2}}}$ 
	with $\sum k_a > m-1$ from $\hat{\mathcal{H}}^{\shtile,-}$ or $\sum k_a > m-2$ from $\hat{\mathcal{H}}^{\shhtile,-}$.
	
	The above four Hamiltonians by the definition have "sum over boxes" property, however we observe that in the expansion in terms of super-Schurs $\mathcal{S}_{\lambda}$ and conjugates $\hat{\mathcal{S}}_{\mu}$ in the spirit of \eqref{single-hook expansion sSchurs Ham} enters diagrams with two and more hooks. We leave this intriguing observation for further investigation elsewhere.

	According to the general reasoning from Section \ref{Schur} we can construct a set of commuting Hamiltonians. We provide the following example, that is based on the Pierri rules for super-Macdonald polynomials:
	\begin{equation}
		\theta_1 \cdot \mathcal{M}^{q,t}_{\lambda} = \sum_{\shtile \in \text{Add}(\lambda)} C_{\lambda, \lambda + \shtile}^{q,t} \cdot \mathcal{M}^{q,t}_{\lambda + \shtile} 
	\end{equation}
	\begin{equation}
		\frac{\p}{\p \theta_1} \ \mathcal{M}^{q,t}_{\lambda} = \sum_{\shtile \in \text{Rem}(\lambda)} C_{\lambda, \lambda - \shtile}^{q,t} \cdot \mathcal{M}^{q,t}_{\lambda - \shtile} 
	\end{equation}
	where coefficients $C_{\lambda, \lambda + \shtile}^{q,t}$ and $C_{\lambda, \lambda - \shtile}^{q,t}$ are rational functions of parameters $q,t$.
	Then we define the following auxiliary operators:
	\begin{equation}
		\hat{E}^{\shtile}_{k \pm 1} = \Big[ \hat{\mathcal{H}}^{\shtile,\pm}, \hat{E}^{\shtile}_{k} \Big]
	\end{equation}
	\begin{equation}
		\hat{F}^{\shtile}_{k \pm 1} = -\Big[ \hat{\mathcal{H}}^{\shtile,\pm}, \hat{F}^{\shtile}_{k} \Big]
	\end{equation}
	These two relations allow increase and decrease index $k$ to define operators for any intger values of $k$ using $\hat{E}^{\shtile}_0 = \theta_1$ and  $\hat{F}^{\shtile}_0 = \frac{\p}{\p \theta_1}$ as the initial conditions of this recursive procedure. Then the commuting set of Hamiltonians read:
	\begin{equation}
		\hat{\mathcal{H}^{\prime}}_{a+b} := \Big\{ \hat{E}^{\shtile}_a, \hat{F}^{\shtile}_b \Big\}
	\end{equation}
	\begin{equation}
		\Big[ \hat{\mathcal{H}^{\prime}}_{a}, \hat{\mathcal{H}^{\prime}}_{b} \Big] = 0
	\end{equation}
	
	We consider here only one possible commuting family of super-Hamiltonians that is related to $\hat{\mathcal{H}}^{\shtile,+}$ and $\hat{\mathcal{H}}^{\shtile,-}$. However, we need to construct a pair of operators that add and remove semi-boxes $\shhtile$ for constructing the remaining commuting family, related to $\hat{\mathcal{H}}^{\shhtile,+}$ and $\hat{\mathcal{H}}^{\shhtile,-}$. We leave it for future research. Of course, these two commuting families commute with each other.

	\section{Conclusion
		\label{conc}}
	
	This paper is devoted to the analysis of Hamiltonians for polynomials of the Schur-Macdonald family and their supersymmetric extensions. These polynomials are naturally enumerated by Young-like diagrams. 
	We argue that the ``sum over boxes'' property of eigenvalues equipped with the Pierri rule for adding/removing boxes provides integrability -- commutativity of an infinite set of Hamiltonians. 
	
	In the case of Schur and super-Schur polynomials each box is weighted with content $\omega = j - i$ that is given by a difference of horizontal and vertical coordinates of the box. 
	Furthermore for the Macdonald case each box is weighted with exponentiated weights $\omega^{q,t} = q^{2j} t^{-2i}$ where $q,t$ are free parameters. 
	Therefore our main idea in constructing super-Hamiltonians is to enforce the special form of eigenvalues: for an eigen function (i.e. a super-Macdonald polynomial) that is labeled by the super-Young diagram the corresponding eigenvalue is given by a sum over all semi-boxes of these diagram with respective exponent weights \eqref{def sup eigval}. 
	
	Super-Hamiltonians are labeled by two indices. 
	The first index labels two different types of semi-boxes $\shtile$ and $\shhtile$ according to the ``sum over boxes'' property of corresponding eigenvalues. 
	The second index encodes positive or negative powers of free parameters $q,t$ in eigenvalues. 
	Therefore the total number of independent super-Hamiltonians is equal to four. 
	However, only two (e.g. two positive) super-Hamiltonians are sufficient to compute all the super-Macdonald polynomials as eigen functions with distinct eigen values.

	There are plenty of details to comment on in this approach -- and each one raises an intriguing question:
	\begin{enumerate}
		\item There is spontaneous breaking of symmetry $q\longrightarrow 1/q, \ t\longrightarrow 1/t$. 
		Macdonald polynomials are invariant with respect to this transformation, whereas Ruijsenaars Hamiltonians are not (this is true about the basic ones and continues to hold for the higher Hamiltonians, see \cite{Mironov:2019uoy}). Surprisingly, this property disappears after super-generalization -- super-Macdonalds are no longer invariant.
		\item This symmetry violation leads to an occurrence of extra Hamiltonians --
		what could be partly related to the apparent difference between Yangians and DIM: there are commuting {\it rays} in the former case. 
		These rays are promoted to entire Heisenberg {\it lines} in the latter \cite{Mironov:2020pcd}.
		\item As it was observed in a variety of examples the ``sum over boxes'' property of eigenvalues is somehow related to a single-hook expansion of the Hamiltonian in terms of eigen functions and their conjugates \cite{Mironov:2019mah}. 
		However, this connection is lost in case of the Macdonald super-Hamiltonians, since they do not have a naive single-hook expansion in terms of super-Schur polynomials.
		\item The projections can be implemented in super-Hamiltonians with the help of
		additional Grassmannian variables -- with their help the formulas acquire almost the same form
		as the Ruijsenaars finite-difference operators (vertex-form operators).
	\end{enumerate}
	
	\section*{Acknowledgments}
	
	The work was partially funded within the state assignment of the Institute for Information Transmission Problems of RAS. Our work is partly supported by the grants of the Foundation for the Advancement of Theoretical Physics and Mathematics ``BASIS'' (A.M. and N.T.).

\bibliographystyle{utphys}
\bibliography{biblio}

\end{document}